# Beam transport experiment with a new kicker control system on the HIRFL-CSR[*]

WANG Yan-Yu(王彦瑜)[1]    ZHOU De-Tai(周德泰)[1]    LUO Jin-Fu(罗金富)[1,2]    ZHANG Jian-Chuan(张建川)[1]    ZHOU Wen-Xiong(周文雄)[1,2,3]    NI Fa-Fu(倪发福)[1,2]    YIN Jun(尹俊)[1,2],    YIN Jia (尹佳)[1]    YUAN You-Jin(原有进)[1]    SHANG-GUAN Jin-Bin(上官靖斌)[1]

（1 Institute of Modern Physics, Chinese Academy of Sciences, Lanzhou, 730000, China
（2 University of the Chinese Academy of Sciences, Beijing, 100039, China
（3 College of Power Engineering, Chongqing University, Chongqing, 400044, China;

**Abstract:** The kicker control system was used for beam extraction and injection between two cooling storage rings (CSRs) at the Heavy Ion Research Facility in Lanzhou (HIRFL). To meet the requirements of special physics experiments, the kicker controller was upgraded. The new controller was designed based on ARM+DSP+FPGA technology and monolithic circuit architecture, which can achieve a precision time delay of 2.5 ns. In September 2014, the new kicker control system was installed in the kicker field, and the test experiment using the system was completed. In addition, a pre-trigger signal was provided by the controller, which was designed to synchronize the beam diagnostic system and physics experiments. Experimental results indicate that the phenomena of "missed kick" and "inefficient kick" were not observed, and the multichannel trigger signals' delay could be adjusted individually for kick power supplies in digitization; thus, the beam transport efficiency was improved compared with that of the original system. The fast extraction and injection experiment was successfully completed based on the new kicker control systems for HIRFL-CSR.

## 1 Introduction

The beam extraction of CSRm and the injection of CSRe are two key components of HIRFL-CSR. The time control precision of the control system affects the efficiency of beam extraction and injection directly. Therefore, the kicker system plays an important role in ring-like accelerators. However, the former kicker control system cannot satisfy the new requirements of special physics experiments because the time control precision of the kick power supplies can only reach 5 ns, which is not sufficiently high[1].

The position information of a bunch cannot be provided directly due to a lack of a bunch phase probe in the HIRFL-CSR. The precise extraction and injection time cannot be accurately determined using programmable logic controllers (PLCs) because the work response time for PLCs cannot reach the order of nanoseconds. At the same time, due to other requirements of new physics experiments, we upgraded the beam kicker control system to improve the control precision and flexibility. The system can now detect the beam phase through the radio frequency (RF) signal indirectly, and the regulation precision can reach 2.5 ns. The schematic layout of the system is shown in Fig. 1. The new system has a monolithic circuit architecture and a faster response time and involves easier maintenance compared to other injection/extraction systems. For example, in the injection and extraction control system of the SSRF, a trigger timing pulse delay distributor is used[2]. The hardware environment is mainly based on the PLC and advanced RISC machine (ARM). The WinCC and the Experimental Physics and Industrial Control System (EPICS) were selected as the software platforms[3]. Similarly, for the BEPCII injection control system, the EPICS control structure is used as the control software, and the PLCs are used as the control hardware. Data and other information are exchanged between the PLC control layer and the EPICS control layer through a VME power supply controller developed by American SNS Laboratory[4]. The advantages of our system are as follows:

1) The hardware and software are designed according to the requirements of the synchrotron system.
2) The friendly interface is designed based on configuration parameters and web communication.
3) The system exhibits a good performance-to-price ratio.
4) The trigger signals, used to control the 6 CSRm kicker power supplies, are delayed through high precise counters instead of monostable circuits. Thus, the unstability and time drift of the signals are better than the old system.
5) Reached a good time resolution.

[*]Supported by National Natural Science Foundation of China（U1232123）
E-mail: yanyu@impcas.ac.cn

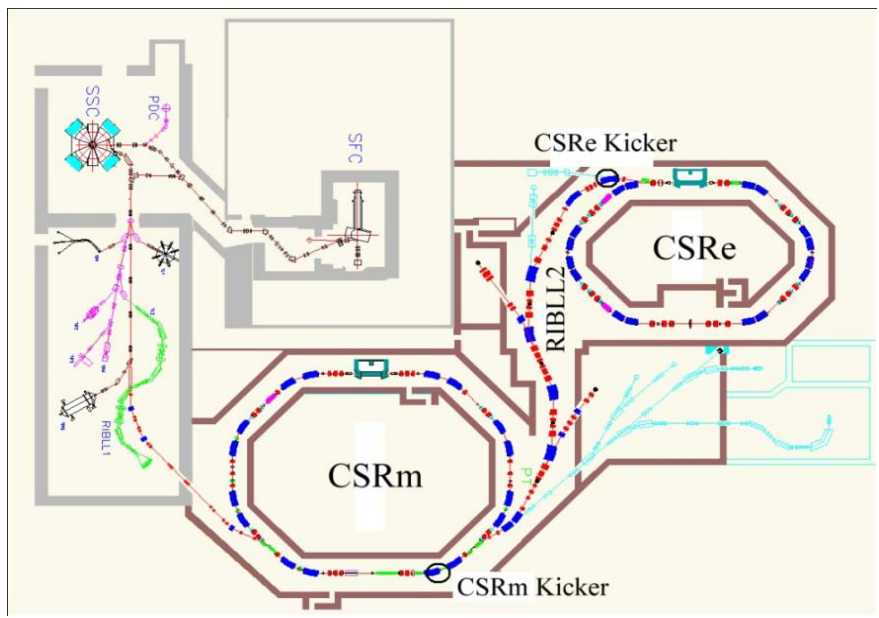

Fig. 1. The schematic layout of the system

For the hardware design, we used a monolithic circuit architecture incorporated with ARM, digital signal processing (DSP) and field programmable gate array (FPGA). The hardware characteristics are presented in table 1. The FPGA chip was upgraded to Cyclone III (400 MHz), which can raise the time control precision of the new control system to 2.5 ns. In addition, the precision can meet the requirements of special physics experiments. Fig. 2 shows the 2.5-ns time precision measured using an oscilloscope.

| IC | original | upgraded |
| --- | --- | --- |
| ARM | ARM7, 50 MHz<br>uClinux | ARM11n, 667 MHz<br>**Linux** |
| FPGA | ACEX1K50, 200 MHz<br>2880 LEs, 40K RAM bits,<br>186 I/O | CycloneIII, **400 MHz**<br>24,624 LEs, 594K RAM bits,<br>215 I/O |
| DSP | DSP5402, 100 MHz | DSPC6713B, 200 MHz, **float** |

Table 1 The hardware characteristics of the former and upgraded control systems

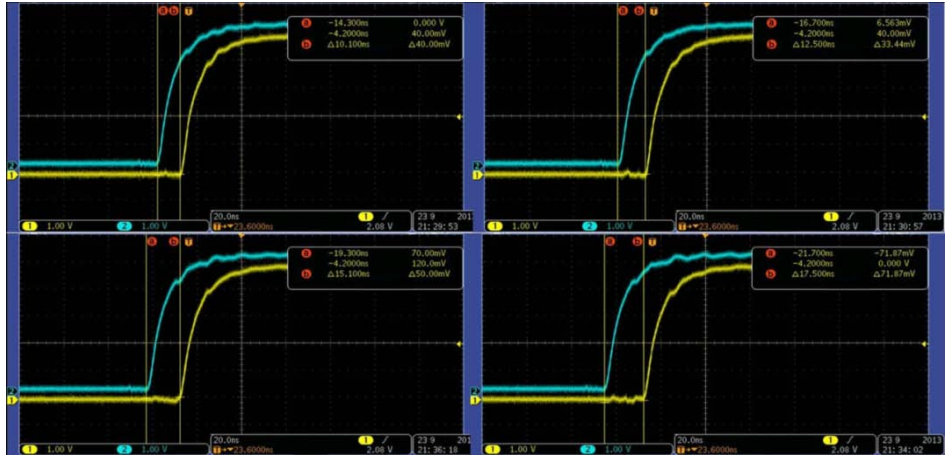

Fig. 2. 2.5-ns time precision

In the human-machine interactive interface (HMI) design process, we used the popular B/S architecture; the Boa web server was transplanted into the embedded Linux operation system in the ARM11, which is used to provide client browser access. The HMI interface of the client was designed using HTML+ CSS +JavaScript and can configure the hardware platform conveniently.

## 2  Experimental conditions

In the HIRFL-CSR, an ion beam is first accumulated in the CSRm, cooled using an electron beam and then accelerated. After the beam attains the required energy, the kicker system is triggered to extract the beam bunches from the CSRm and inject these beam bunches over the RIBLL2 into the CSRe or other experimental terminals for physics experiments.

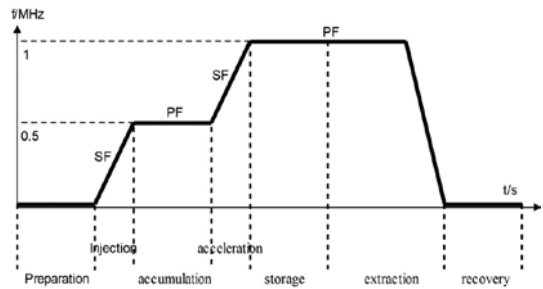

Fig. 3. Working cycle of CSRm (SF denotes the sweep frequency, and PF denotes the point frequency).

The HIRFL-CSR operation mode "ECR+SFC+CSRm+CSRe" was used in the experiment conducted in this study. A $C^{3+}$ beam was generated by ECR for this experiment. The beam energy reached 366 MeV in the CSRm. As shown in Fig. 3, the entire process consists of seven steps: preparation, beam injection (from RIBLL1 to CSRm), beam accumulation, beam acceleration, beam storage, beam extraction and state recovery[5]. Each step is assigned an event code, which is used as a trigger flag to control the operation of corresponding devices at different steps. The charging and discharging trigger signals for kick power supplies are sent when the kicker control system receives the event code for beam extraction and the RF signal. To complete the entire control process in an efficient manner, two characteristic times must be accurately determined: the kick time of the CSRm (kicks the particle beam bunches out) and CSRe (kicks the particle beam bunches in) and the delay time, *i.e.*, the signal transmission time from the CSRm to CSRe. However, because there is no beam phase probe device at the CSRm, it is not possible to directly detect the positions of the beam bunches. Thus, we inferred the position by measuring the phase of the RF signal. The RF signal is a point frequency signal that occurs during the extraction stage; the beam bunch position can therefore be detected indirectly for the synchronization of the phase of the RF signal and the bunch. Thus, the time of beam extraction must be accurately synchronized with the phase of the RF signal. As is known, six power supplies are used to drive a kicker magnet for the extraction of beam bunches in the CSRm, and the other four power supplies are used to drive another kicker magnet for the injection of beam bunches into the CSRe. The distance between the two sets of kick power supplies is approximately 200m, and the distance between the two sets of kicker magnets is approximately 120m[5]. The flight time of the beam bunches must be taken into account at these distances when the discharge delay time is adjusted for both the CSRm and CSRe. Therefore, the trigger signal for CSRe should be sent early. The entire working cycle of CSRm can be completed within 20 to 35s[6].

## 3 Experimental methods

The former kicker control system consists of four PCB boards: a communication board, control board, signal delaying board and reference voltage setting board. Backplane buses are used to connect the four circuit boards. The upgraded controller uses a monolithic circuit architecture for all functions; however, the reference voltage setting function can be integrated into one PCB board. This type of structure can effectively reduce the delay time of the hardware system and the extra interference signal caused by an excessive number of external connection signals. In this respect, to ensure the normal operation of the entire system, all control signals must be reconnected to the corresponding equipment. Related signals include the kicker event code, RF signal, charging and discharging trigger signals for the kick power supplies, event trigger and synchronous signal for beam diagnostics and physics experiments. These signals must be connected to the controller correctly. Individual optical fibers are provided for each power supply in the new system, which can adjust the delay time individually.

The new system can accurately acquire the position of the beam bunches (mentioned in section 2) and generate the control signals. The frequency signal is compared tens of thousands of times inside the FPGA to determine the vertex frequencies. First, a pre-trigger signal is provided by the controller for the beam diagnostic system and physics experiments. Then, the controller can send a signal to drive the kick power supplies to complete the beam extraction procedure. At the same time, this controller can synchronously send a discharging signal to trigger the injection kick power supplies of CSRe, and the delay time between the CSRm and CSRe discharging trigger signals can be adjusted to match the flight time of bunches over RIBLL2. The hardware platform is shown in Fig. 4.

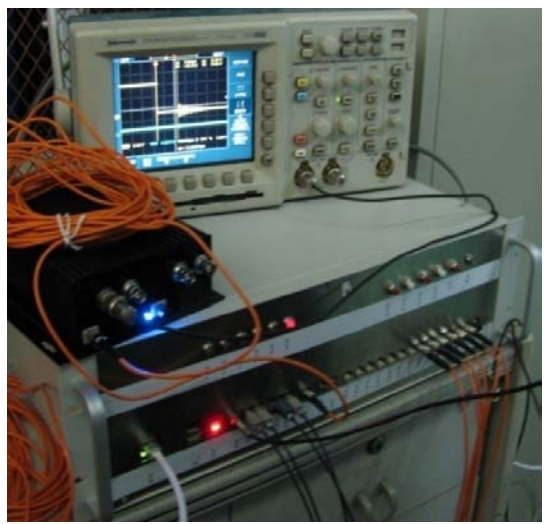

Fig. 4. The hardware platform

In this control system, the software can acquire data from the controller through the TCP/IP protocol. The human-machine interactive interface (HMI) is shown in Fig. 5. The browser can access the kicker controller through a unique IP address. The design of the software is based on the B/S architecture, and the Boa web server is transplanted into the embedded Linux operation system in the ARM11 to provide browser access.

HTML/CSS/JavaScript technologies are used to develop the client user interface for the operator to configure the hardware platform.

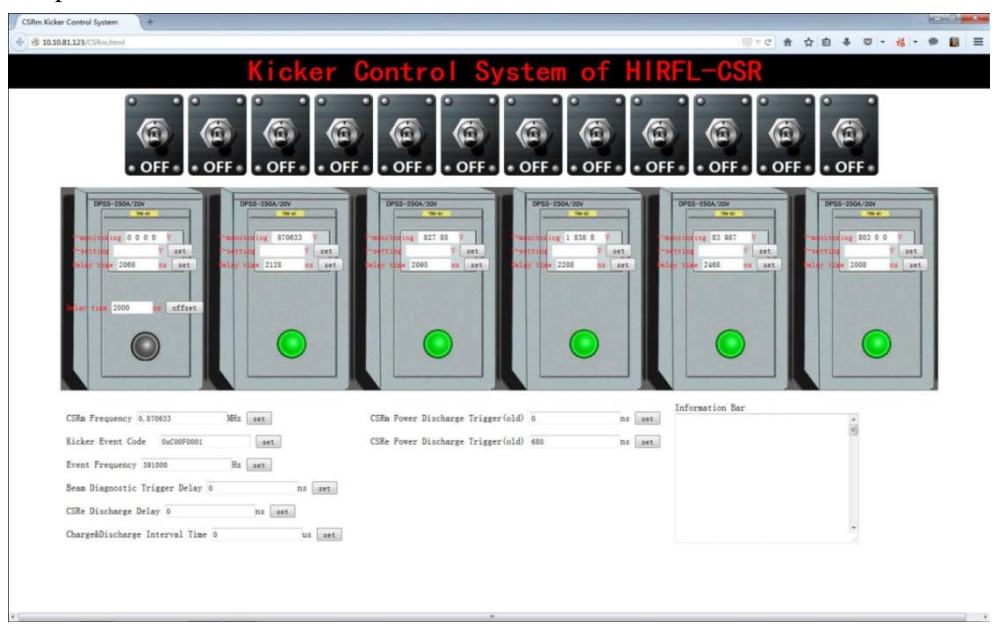

Fig. 5. The user-machine interface

## 4 Parameter settings

The upgraded controller delivers the charging and discharging trigger signals to six kick power supplies on the CSRm when receiving the kicker event code and capturing the phase of the RF signal. At the same time, this controller can synchronously send a signal to the CSRe to trigger the injection function and a pre-trigger signal for beam diagnostics and physics experiments. In addition, the pre-trigger signal is provided by the controller to meet the requirements of the beam diagnostic system and physics experiments. All delay times can be adjusted individually, and the interval time precision that can be achieved is up to 2.5 ns in the digital delayer.

The controller parameters are set as follows.

1) The event code is 0xC00F0001.
2) The extraction vertex frequency is set to 0.870633 MHz.
3) The six kick power supplies used to drive kicker magnets for the extraction of beam bunches on the CSRm side are set to 19.2 kV.
4) The four power supplies used to drive kicker magnets for the injection of beam bunches on the CSRe side are set to 32.8 kV.
5) The discharge times of the six power supplies of the CSRm are set to 1500 ns which compare to CSRe.
6) The delay times of the four power supplies of CSRe are synchronized to 680 ns.
7) The pre-trigger delay time of beam diagnostics is set to 0 ns.
8) The physics experiments pre-trigger delay time is also set to 0 ns. This delay time takes precedence over the discharge trigger time of approximately 3.6 µs.

After setting the parameters described above and verifying that the kicker event code and the phase of the RF signal are normal, experiments are ready to begin. The waveform of discharge for extraction captured by an oscilloscope is shown in Fig. 6.

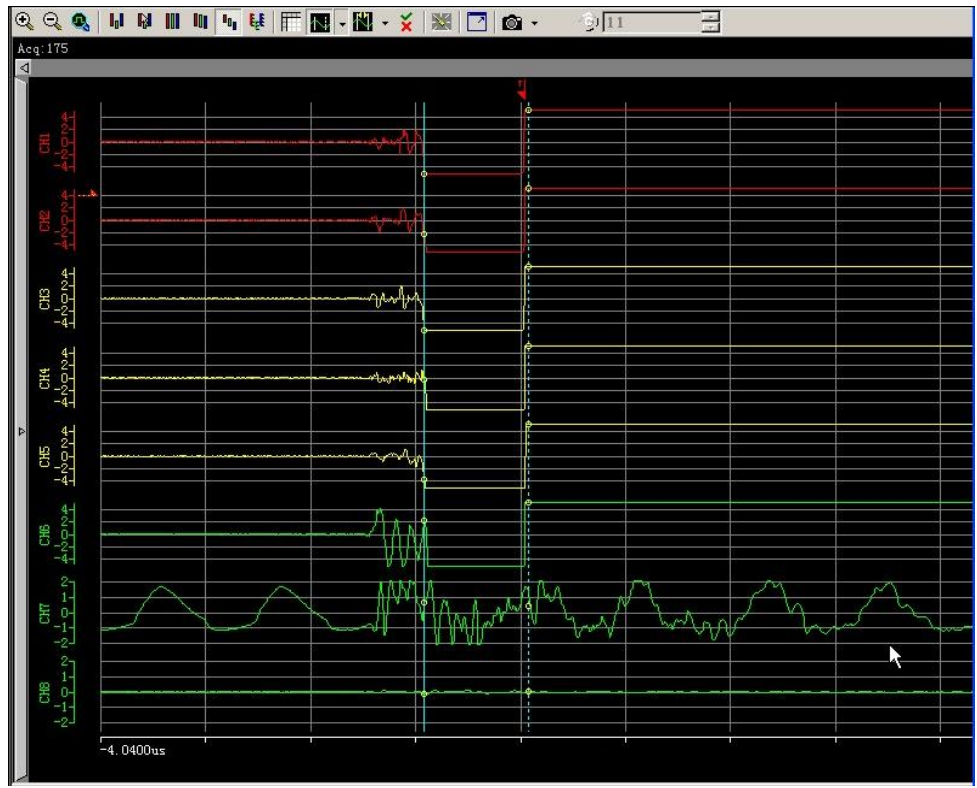

Fig. 6. Waveform of discharge for extraction

## 5 Experimental conclusions

The new kicker system was used successfully to extract the beam bunches from the CSRm and inject them over the RIBLL2 in the CSRe in this study. Figs. 7 and 8 show beam extraction from the CSRm and beam injection into the CSRe, respectively. The upgraded system has six specific advantages. First, the control precision of the kicker system is upgraded to 2.5 ns based on the newly designed ARM+DSP+FPGA technology. Therefore, the phase capture accuracy and capture rate are increased, and the phenomena of "leaky kick" and "kick empty" are not observed. Second, the upgraded controller uses a monolithic circuit architecture. A single board can complete all functions except for the reference voltage setting function. The system eliminates various types of interference and protects the devices from harsh conditions. Thus, the stability of the system is improved, and the complexity of operation and maintenance is reduced. Third, the discharging time delay of each kick power supply can be independently adjusted, which can satisfy the requirements of different experiments and can provide considerable flexibility. Moreover, time motility is avoided, and the placement data show good repeatability. Fourth, a glass fiber is used to reduce the delay time caused by the repeater relative to that of the old POF fiber, which improves the synchronous control function of the two kicker systems of the CSRm and the CSRe. Fifth, the development of the software is based on the B/S architecture, and the Boa web server is transplanted into the embedded Linux in ARM11 to provide browser access. Finally, the controller can send a stable pre-trigger signal for beam diagnostics and physics experiments.

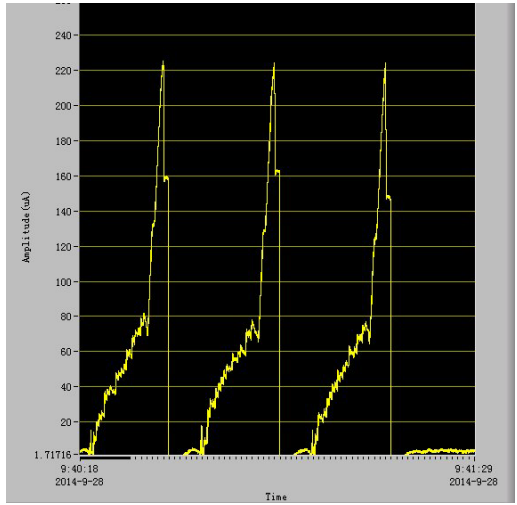

Fig. 7. Beam extraction from CSRm

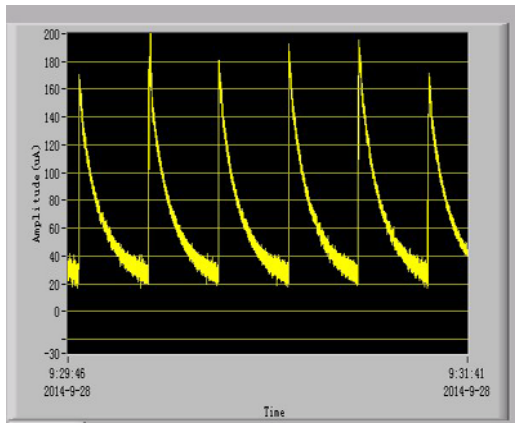

Fig. 8. Beam injection into CSRe

## 6 Further plans

In this upgraded system, the position of beam bunches can be determined by measuring the phase of the radio frequency (RF) signal in real time because the beam bunches are synchronized with the RF signal. Based on the experimental results obtained in this study, the hardware and software design fully meet the requirements of kicker control for CSR and new physics experiments. In addition, the new pre-trigger signal can satisfy the requirements of the beam diagnostic system. The results show that the new controller can solve the problems associated with the former system. The original operation kicker controller, which was also designed by our group and had been running since March 2007, can be replaced by the upgraded kicker controller. Thus, we plan to replace the former kicker control system for CSR during the summer of 2015. Based on the test results obtained for the new controller in this study, several improvements will be made. First, the software and HMI should be optimized and improved, and a friendlier interface will be provided to users. Second, the kick power supply setting and read-back functions will be joined. The controller can communicate with a self-developed module ICP400 to control the reference voltage setting through the RS485 interface, and the accuracy of the reference voltage setting can reach as low as 0.1%, which is better than that of the former system (5%). Third, remote switching and status monitoring functions for the kick power supplies will be added. Finally, the new kicker control system will be run to improve the efficiency of beam extraction and beam injection of the HIRFL-CSR.

## 7 Author information

WANG Yan-Yu(王彦瑜) Institute of Modern Physics, CAS, Nanchang Rd. 509, Lanzhou, China 730000, 0931-4969357, yanyu@impcas.ac.cn

ZHOU De-Tai(周德泰) Institute of Modern Physics, CAS, Nanchang Rd. 509, Lanzhou, China 730000, 0931-4969769, zhoudet@impcas.ac.cn

LUO Jin-Fu(罗金富) Institute of Modern Physics, CAS, Nanchang Rd. 509, Lanzhou, China 730000, 0931-4969768, luojinfu@impcas.ac.cn

ZHANG Jian-Chuan(张建川) Institute of Modern Physics, CAS, Nanchang Rd. 509, Lanzhou, China 730000, 0931-4969769, zhangjc@impcas.ac.cn

ZHOU Wen-Xiong(周文雄) College of Power Engineering, Chongqing University, Shazheng Rd. 174, Chongqing, China, 400044, 023-65102473, zhouwenxiong@cqu.edu.cn

NI Fa-Fu(倪发福) Institute of Modern Physics, CAS, Nanchang Rd. 509, Lanzhou, China 730000, 0931-4969768, nifafu@impcas.ac.cn

YIN Jun(尹俊) Institute of Modern Physics, CAS, Nanchang Rd. 509, Lanzhou, China 730000, 0931-4969768, yinjun@impcas.ac.cn

YIN Jia (尹佳) Institute of Modern Physics, CAS, Nanchang Rd. 509, Lanzhou, China 730000, 0931-4969769, yinjia@impcas.ac.cn

YUAN You-Jin(原有进) Institute of Modern Physics, CAS, Nanchang Rd. 509, Lanzhou, China 730000, 0931-4969509, yuanyj@impcas.ac.cn

SHANG-GUAN Jin-Bin(上官靖斌) Institute of Modern Physics, CAS, Nanchang Rd. 509, Lanzhou, China 730000, 0931-4969018, sgjb@impcas.ac.cn